\documentclass[rnote]{aa} 
\usepackage{graphicx}
\usepackage{txfonts}
\begin{document}
   \title{Cygnus~X-3 transition from the ultrasoft to the hard
   state\thanks{Based on observations with INTEGRAL, an ESA project
   with instruments and science data centre funded by ESA member
   states (especially the PI countries: Denmark, France, Germany,
   Italy, Switzerland, Spain), Czech Republic and Poland, and with the
   participation of Russia and the USA}
 }


   \author{V. Beckmann
          \inst{1,2,3},
           S. Soldi\inst{1,2},
           G. B\'elanger\inst{4}, 
           S. Brandt\inst{5},
           M. D. Caballero-Garc\'{\i}a\inst{6}, 
           G. De Cesare\inst{7,8,9}, 
	   N. Gehrels\inst{10}, 
           \mbox{S. Grebenev\inst{11}},
           O. Vilhu\inst{12},
           A. von Kienlin\inst{13},
           \and
           {\mbox T.J.-L. Courvoisier\inst{1,2}}
           }

   \offprints{V. Beckmann}

   \institute{INTEGRAL Science Data Centre, Chemin d'\'Ecogia 16, 1290 Versoix, Switzerland\\
\email{Volker.Beckmann@obs.unige.ch}
     \and
   Observatoire Astronomique de l'Universit\'e de Gen\`eve, Chemin des Maillettes 51, 1290 Sauverny, Switzerland
     \and
   CSST, University of Maryland Baltimore County, 1000 Hilltop Circle,
   Baltimore, MD 21250, USA
   \and
    INTEGRAL Science Operations Centre,
	European Space Astronomy Centre (ESAC),
	Apartado 50727,
	28080 Madrid, Spain
   \and
   Danish National Space Centre, Technical University of Denmark, Juliane Maries Vej 30, 2100 Copenhagen, Denmark
 \and
 Laboratorio de Astrof\'isica Espacial y F\'isica Fundamental (LAEFF-INTA), POB 50727, 28080 Madrid
   \and
      INAF-Istituto di Astrofisica Spaziale e Fisica Cosmica di Roma, via Fosso del Cavaliere 100, I-00133 Roma, Italy
    \and
Dipartimento di Astronomia, Universita' degli Studi di Bologna, Via
Ranzani 1, I40127 Bologna, Italy
\and
 Centre d'Etude Spatiale des Rayonnements, CNRS/UPS, B.P. 4346, 31028
 Toulouse Cedex 4, France
\and
 Astrophysics Science Division, NASA Goddard Space Flight Center, Code
 661, MD 20771, USA
 \and
  Space Research Institute, Russian Academy of Sciences, Profsoyuznaya 84/32, 117997 Moscow, Russia
 \and
 Observatory, PO Box 14, University of Helsinki, FIN-00014 University of Helsinki, Finland
 \and
 Max-Planck-Institut f\"ur extraterrestrische Physik,
 Gie{\ss}enbachstra{\ss}e, 85748 Garching, Germany
\\ 
}
             
\authorrunning{Beckmann et al.}

   \date{Received June 26, 2007; accepted August 3, 2007}

 
  \abstract
   {}
   {The nature of Cygnus~X-3 is still not understood well. This binary
   system might host a black hole or a neutron star. Recent observations by INTEGRAL
   have shown that Cygnus~X-3 was again in an extremely ultrasoft state. Here we present our
   analysis of the transition from the ultrasoft state, dominated by
   blackbody radiation at soft X-rays plus non-thermal emission in the hard X-rays, to the low hard state.}
   {INTEGRAL observed Cyg X-3 six times during three weeks in late May and early June
   2007. Data from IBIS/ISGRI and JEM-X1 were analysed to show the
   spectral transition.}
   {During the ultrasoft state, the soft X-ray spectrum is well-described by an absorbed ($N_H = 1.5 \times 10^{22} \rm \,
   cm^{-2}$) black body model, whereas the X-ray spectrum above
   20 keV
   appears to be extremely low and hard ($\Gamma \simeq 1.7$). During
   the transition, the radio flux rises to a level of $> 1 \rm \, Jy$,
   and the soft X-ray emission drops by a factor of $\sim 3$, while the
   hard X-ray emission rises by a factor of $\sim 14$ and becomes steeper (up to $\Gamma = 4$).}
   {The ultrasoft state apparently precedes the emission of a
   jet, which is apparent in the radio and hard X-ray domain.}


   \maketitle


\section{Introduction}

The binary system Cygnus X-3 is one of the brightest objects in the
X-ray sky. It was discovered by one of the first X-ray rocket
experiments (Giacconi et al. 1967). Although one of the longest
known and brightest X-ray sources, its nature is still not understood very well. The object is bright throughout the electromagnetic
spectrum (e.g. McCollough et al. 1999) and is located in the
Galactic plane at a distance of 9 kpc (Predehl et al. 2000). In this system, a donor star is orbiting a compact
object in a close orbit. The nature of
the compact object is still under debate, and it could be a black hole,
an X-ray pulsar, or a neutron star with a low magnetic field. {\it
  Chandra} observations indicate that the masses
of the donor star and the compact object are $M \le 7.3 M_\odot$ and
$M \le 3.6 M_\odot$, respectively (Stark \& Saia 2003). Strong absorption features are observed
throughout the spectrum, which leads to the assumption that the system
is embedded in a dense wind coming from the donor, presumably a
massive, nitrogen-rich Wolf-Rayet star with huge mass loss (\cite{CygX3WR}).
The light curve of the system shows a 4.8hr quasi-sinosoidal
modulation, present both in X-rays and infrared (\cite{Goldoni03}).

Cygnus~X-3 is also a bright radio source, with a quiescent flux $< 100
\rm \, mJy$, a flux of 0.1 to 1 Jy during minor flares, and more than
1 Jy in strong flares (McCollough et al. 1999). The radio emission
during strong flares appears to be correlated to the hard X-ray emission, but
anticorrelated with the soft X-rays, whereas the situation is inverted
during quiescence (\cite{McCo}). As the source shows radio-jet
like structures during flares (Miller-Jones et al. 2004) and
quasi-periodic oscillations (van der Klis \& Jansen 1985), Cygnus~X-3 might
be a microquasar.

Cygnus~X-3 was studied by {\it INTEGRAL} (Winkler et al. 2003)
early during the mission (Vilhu et al. 2003). Lately, Cyg X-3 appeared
again to be in an ultrasoft state with strong soft X-ray but weak
hard X-ray emission (Soldi et al. 2007). Here we report on our 
detailed analysis of the ultrasoft state and its transition to the
``normal'' low hard state. 

\section{Data analysis}

INTEGRAL (\cite{INTEGRAL}) observed \mbox{Cygnus~X-3} several times after it reached the
ultrasoft state in May 2007 (\cite{ATel2007}). 
Data from the imager IBIS/ISGRI (\cite{ISGRI}) and the X-ray monitor JEM-X1 (\cite{JEMX})
were analysed using the Offline Standard Analysis package OSA 6
provided by the ISDC (\cite{ISDC}). 
Table~1 gives a list of the observations performed.

\begin{table*}
\caption[]{INTEGRAL observation log}
\label{obslog}
\begin{tabular}{lllll}
Revolution & Start time & Stop time & ISGRI exposure & significance\\
           & [U.T.]     &    [U.T.] &    [sec]       & $[\sigma]$\\
\hline
562     &   2007-05-21T09:45 & 2007-05-21T23:37 & 9557   &        5.2\\
563     &   2007-05-25T07:35 & 2007-05-26T21:43 & 51237   &        3.1\\ 
564     &   2007-05-27T08:53 & 2007-05-29T21:25 & 64319   &        5.8\\
567     &   2007-06-05T08:18 & 2007-06-07T21:35 &  39750  & 72.3 \\
568     &   2007-06-08T08:08 & 2007-06-10T20:45 &  21750  & 56.9 \\
569     &   2007-06-11T07:57 & 2007-06-12T02:40 &  21201  & 54.6 \\
\end{tabular}
\end{table*}

\begin{figure}
\centering
\includegraphics[width=8.5cm]{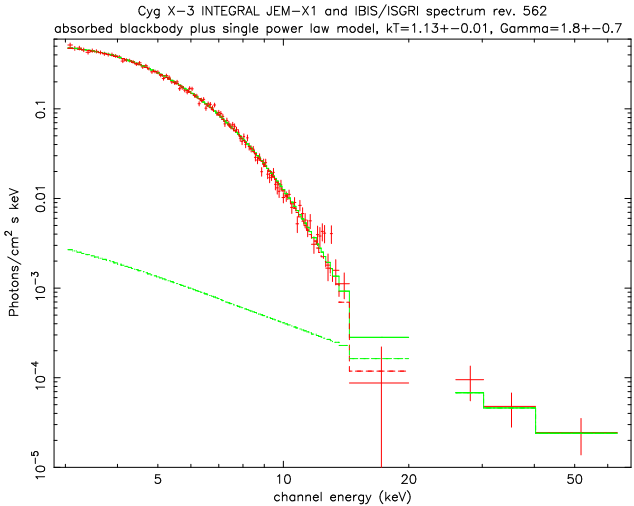}
\caption{Revolution 562 data. The combined spectral plot uses the INTEGRAL
  JEM-X1 and IBIS/ISGRI data of Cygnus~X-3 ($\chi^{2}_\nu =
  1.26$).}
              \label{fig:ultrasoft}
\end{figure}

JEM-X1 covered the position of Cyg X-3 only during revolution 562 and
detected the source within 6945s with a 175 sigma significance. The
combined JEM-X1 and ISGRI spectrum in rev. 562 allows a detailed
spectral modelling. The best-fit results are achieved for an
absorbed ($N_H = (1.5 \pm 0.4) \times 10^{22} \rm \, cm^{-2}$)
blackbody ($kT = 1.13 \pm 0.01 \rm \, keV$) plus power law with $\Gamma =
1.8 \pm 0.7$ ($1 \sigma$ errors). The luminosity over the 2--60 keV range is $L_X = 1.1
\times 10^{38} \rm \, erg \, s^{-1}$, assuming a distance of 9 kpc
(Predehl et al. 2000). With a flux in the 2--10 keV band of $f_X = 1.1
\times  10^{-8} \rm \, erg \, cm^{-2} \, s^{-1}$ and only $f_X =
10^{-10} \rm \, erg \, cm^{-2} \, s^{-1}$ at 20-60 keV, Cyg X-3 was
clearly in an ultrasoft state during this observation. The spectrum of
revolution 562 is shown in Fig.~\ref{fig:ultrasoft}. 

\begin{figure}
\centering
\includegraphics[width=10cm]{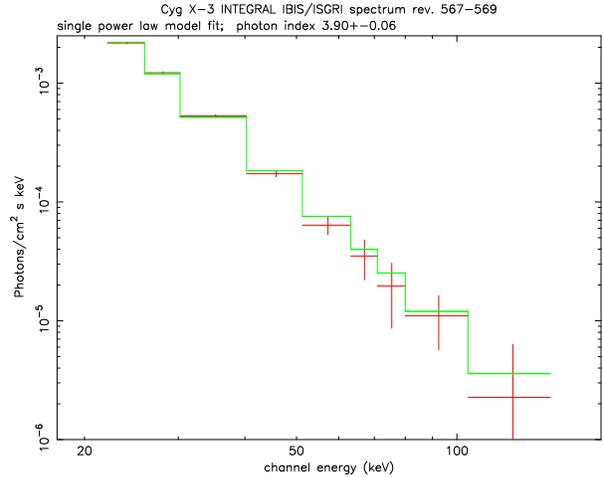}
\caption{Revolution 567-569 data. The spectral plot uses the INTEGRAL IBIS/ISGRI data of Cygnus~X-3.}
              \label{fig:lowhard}
\end{figure}

{\it INTEGRAL}
observed the region several times over the following three weeks. The
IBIS/ISGRI data can be fit in all cases with a single power-law model. The
fluxes, photon indices, and luminosities based only on ISGRI data
are given in Table~2. In addition, we list there
the fluxes taken from {\it RXTE}/ASM in the 2--10 keV range. The ASM count
rates, based on
the quick-look results provided by the ASM/{\it RXTE} team, have been
converted assuming a Crab-like spectrum where the Crab
has a flux of $f_{2-10 \rm \, keV} = 2.2 \times 10^{-8} \rm \, erg
\, cm^{-2} \, s^{-1}$ (\cite{crabflux}). It has to be taken into
account, though, that the errors of the measurements are likely to be
underestimated for the following reasons, as reported by the instrument
team. A single elevation correction based on the 2--10 keV flux from
the Crab nebula is used. This elevation correction, however, has a
spectral dependence because of increased low-energy absorption in the
thermal shields and counter windows. The sources of a different spectral
shape than the Crab's, then, may well have greater scatter. Second, the
Cygnus region harbours the very bright and variable source Cyg~X-1, which might
contribute to additional systematic error in the ASM data. The results based on the {\it INTEGRAL} data 
revolution 567-569 have been combined as there were no differences
within the error bars. In Fig.\ref{fig:lowhard} we show the IBIS/ISGRI spectrum of
revolution 567-569, which shows that Cygnus~X-3 is indeed in the low hard
state again. The spectrum can be described by a single power-law model with photon index $\Gamma = 3.94 \pm 0.03$.

\begin{table*}
\caption[]{INTEGRAL IBIS/ISGRI results}
\label{intresults}
\begin{tabular}{lllclc}
Revolution & $f_{20- 40 \rm \, keV}$ & $f_{40- 60 \rm \, keV}$  &
           $\Gamma$ & $L_{20 - 60 \rm \, keV}$ & ASM flux (2-10 keV)\\
           & $[\rm \, erg \, cm^{-2} \, s^{-1}]$& $[\rm \, erg \, cm^{-2}
           \, s^{-1}$] & & $[\rm \, erg \, s^{-1}$] & $[\rm \, erg \,
           cm^{-2} \, s^{-1}$]\\ 
\hline
562      &   $5.6\times 10^{-11}$ &  $3.5\times 10^{-11}$&
            $1.9 \pm 0.9$
           &  $8.8\times 10^{35}$ & $(7.22 \pm 0.17) \times 10^{-9}$\\
563      &   $5.5\times 10^{-11}$ & $3.8\times 10^{-11}$&  $1.7 \pm
           1.0$  & $8.9\times 10^{35}$ & $(8.30 \pm 0.17) \times 10^{-9}$\\ 
564      &  $8.6\times 10^{-11}$ & $2.5\times 10^{-11}$& $3.2 \pm 0.6$
           & $1.1\times 10^{36}$ & $(7.73 \pm 0.09) \times 10^{-9}$\\
567-569   & $1.1\times 10^{-9}$ & $2.1\times 10^{-10}$& $3.94 \pm
           0.03$ & $1.3 \times 10^{37}$ & $(3.49 \pm 0.04) \times 10^{-9}$ \\

\end{tabular}
\end{table*}

\section{Discussion and conclusions}

\begin{figure}
\centering
\includegraphics[width=6.7cm,angle=90]{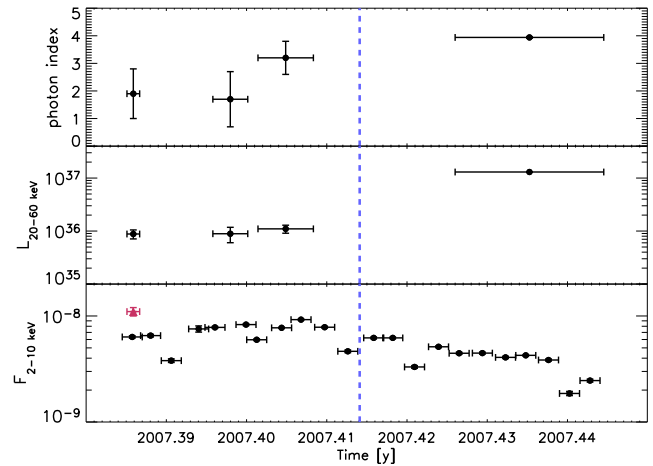}
\caption{The evolution of the photon index $\Gamma$ of the hard X-ray
  component, the luminosity $L_X$ (in $[\rm erg \, s^{-1}]$) in the 20 -- 60 keV energy
  band, and the 2--10 keV flux (in $[\rm erg \, cm^{-2} \, s^{-1}]$) versus time. The soft X-ray flux values
  are taken from {\it RXTE}/ASM except for one JEM-X1 data point,
  which is indicated by the triangle on the left of the bottom
  panel. The dotted line indicates the occurance of the radio flare as
  reported by \cite{Trush07a}.}
              \label{fig:evolution}
\end{figure}

It is apparent that the transition from the ultrasoft state to the low
hard state is
accompanied by a steepening of the  high-energy
component. At the same time, the luminosity of this component has increased
by a factor of $\sim 14$. A radio observation within the transition
phase on 2007-06-01 (Trushkin et al. 2007a) showed strong emission ($>
1$ Jy), which indicates increased jet activity. The behaviour
during this transition from the ultrasoft to low hard states is comparable
to what was reported in January 2007 (Trushkin et al. 2007b). The
ultrasoft state is apparent during spacecraft revolutions 562 and 563,
with an energy output of $E f_E = 0.05 \rm \, keV \, cm^{-2} \,
s^{-1}$ in the 20 -- 60 keV energy band. In revolution 564, we observed
the onset of the transition, with a steeper spectral slope in the hard
X-rays but still comparably low luminosity.
Figure~\ref{fig:evolution} shows the evolution of the spectral slope
and of the luminosity in the 20--60 keV energy band with time.
In revolution 567, the low hard state is reached and the ISGRI
  data can be modelled by a simple power law with $\Gamma = 3.94 \pm
  0.03$. The low hard state has been measured in this energy range
  before, e.g. by {\it RXTE}/HEXTE (Choudhury \& Rao 2002; Choudhury
  et al.
  2002). The data with higher significance allow them to apply a more
  complex model, i.e. Comptonization of seed photons from a thermal
  multi-coloured accretion disk by a thermal Comptonizing plasma cloud
  (CompST; \cite{CompST}) with electron temperature $kT_e
  = 4.9 \pm 0.1 \rm \, keV$ and a
  single power law with $\Gamma  = 2.01 \pm 0.04$. Although the ISGRI data are
  described well by a simple power law with $\Gamma  = 3.9$, we
  applied the more complex model for comparison reasons, which results
  in $kT_e = 5.2 \pm 4.0 \rm \, keV$ and $\Gamma = 3.4 \pm 1.0$. When freezing
  the power law to the value reported by Choudhury \& Rao, the
  electron temperature becomes $kT_e = 8.6 \pm 2.1 \rm \, keV$. Thus, to
  explain the observed ISGRI spectrum in terms of a Comptonization
  component plus a power law, a stronger Comptonization component is
  observed here than by {\it RXTE}/HEXTE in May 1998. It has to be kept in mind,
  though, that the ISGRI data of the low hard state alone are well-fit by a single power law
  and do not require more complex modelling.
 
 The actual transition from the high soft to the low hard state took
place during revolutions 565 and 566 as observed in the radio band.  
As we are missing the X-ray observation during this period, we
cannot say whether the source has undergone an intermediate state
as described in Szostek \& Zdziarski (2004). The {\it RXTE}/ASM data
show, however,  that the transition has to be rather smooth, as the 2--10
keV flux decreases gradually from the ultrasoft state ($f_{2-10 \rm \,
  keV} = 8.3 \times 10^{-9} \rm \, erg \, cm^{-2} \, s^{-1}$) to the low
hard state ($f_{2-10 \rm \,
  keV} = 2.5 \times 10^{-9} \rm \, erg \, cm^{-2} \, s^{-1}$).

Szostek \& Zdziarski (2004) suggest that the transition from the
ultrasoft state to the low hard state is a transition in the hard
X-rays from a jet-dominated phase to thermal Comptonization. In our
observation, this would mean that we detect the emission of the jet in the
ultrasoft state in the
hard X-rays as indicated by $\Gamma \sim 2$. During the transition the jet becomes more and more
diluted by the onset of the strong thermal Comptonization component,
which we see in the steepening and brightening of the hard X-ray spectrum. 
Rajeev et al. (1994) interpreted the transition in terms of an
increasing temperature of the black body component, which is at the
same time decreasing in size and whose Comptonization region
simultaneously is becoming more compact and more opaque. Finally, the hard
X-ray spectrum of the low hard state can be fit by a model of almost
pure Compton reflection (Hjalmarsdotter et al. 2004, 2007).

The true nature of the compact object in Cygnus X-3 still has to be
determined. Hard X-ray observations, as provided by {\it INTEGRAL} and
simultaneous radio observations, are essential in disentangling
the four main components: the absorbing material detectable in soft
X-rays, the thermal (blackbody) component, the jet, and the Comptonization component.

\end{document}